\documentclass[journal=nalefd, manuscript=letter, layout=twocolumn]{achemso}

\usepackage{graphicx}
\usepackage[version=3]{mhchem} 
\usepackage{color}
\usepackage{subfigure}

\title[Nanotwin Networks in Germanium]  {Formation of Nanotwin Networks during High-Temperature Crystallization of Amorphous Germanium}

\author{Luis Sandoval}
\affiliation[Los Alamos National Laboratory]{Los Alamos National Laboratory, Los Alamos, NM 87545}
\author{Celia Reina}
\affiliation[University of Pennsylvania]{University of Pennsylvania, Philadelphia, PA 19104}
\author{Jaime Marian}
\altaffiliation{Formerly at: Physical and Life Sciences Directorate, Lawrence Livermore National Laboratory, Livermore, CA 94551.}
\email{jmarian@ucla.edu}
\affiliation[University of California Los Angeles]
{University of California Los Angeles, Los Angeles, CA 90095}

\abbreviations{Germanium, Crystallization, Molecular dynamics, Twinning, X-ray diffraction}

\begin{document}

\begin{abstract}
Germanium is an extremely important material used for numerous functional applications in many fields of nanotechnology. In this paper, we study the crystallization of amorphous Ge using atomistic simulations of critical nano-metric nuclei at high temperatures. We find that crystallization occurs by the recurrent transfer of atoms via a diffusive process from the amorphous phase into suitably-oriented crystalline layers. We accompany our simulations with a comprehensive thermodynamic and kinetic analysis of the growth process, which explains the energy balance and the interfacial growth velocities governing grain growth. For the $\langle111\rangle$ crystallographic orientation, we find a degenerate atomic rearrangement process, with two zero-energy modes corresponding to a perfect crystalline structure and the formation of a $\Sigma3$ twin boundary. Continued growth in this direction results in the development a twin network, in contrast with all other growth orientations, where the crystal grows defect-free. This particular mechanism of crystallization from amorphous phases is also observed during solid-phase epitaxial growth of $\langle111\rangle$ semiconductor crystals, where growth is restrained to one dimension. We calculate the equivalent X-ray diffraction pattern of the obtained nanotwin networks, providing grounds for experimental validation.
\end{abstract}

\section{\label{sec:intro}Introduction}

Growth of semiconductor crystals from glassy or vapor phases is an extremely important process for many applications in (nano) technology\cite{greene1983review,shay2013ternary}. 
In general, crystallization from a disordered structure is ultimately a diffusive process\cite{jonsson2000theoretical,bourgoin1984solid} and --as such-- strongly temperature dependent. Growth, however, is highly susceptible to the formation of crystal defects, which can be copious, mediated by imperfections, both related to the environmental variables of the physico-chemical treatment \cite{hurle2004brief}, and to structural heterogeneities associated with the substrate, such as impurities, lattice mismatch, etc.\ \cite{drosd1982some}. Defect formation is typically also temperature dependent and thus a compromise must be found to balance reasonable growth rates while keeping acceptably-low defect concentrations. 
Since the early times of solid-phase crystallization, great emphasis has been placed on suppressing the emergence of these imperfections\citep{brown1988theory,dash1959growth}, as many properties
of crystallized systems strongly depend on achieving pristine structures and
a defect-free finish \cite{kohn1958,hwang1995solid}.

Among the panoply of possible defects found during crystal growth, growth twins stand out as one of the more prolific ones due to low formation energies and a variety of possible genesis pathways \cite{donnelly1967,drosd1982some,bulling1956growth}. In this work we are concerned with twinning in diamond-cubic structures, such as Si and Ge, whose growth, solidification, and recrystallization have been studied extensively\cite{drosd1982some,brown1988theory,billig1955growth,cullis1971epitaxial}. It is believed that twinning may appear as a consequence of internal transformations to rotate the crystal structure towards energetically favorable interfacial orientations\cite{bulling1956growth,carstens1968}. However, twinning is also observed during epitaxial recrystallization in Si, Ge, and their alloys, particularly along the $\langle111\rangle$ growth direction\cite{darby2013}.
The chief difficulty in studying twin nucleation in these systems
is that it is generally impossible to see them form \emph{in situ}. Several authors have hypothesized twins form at a very early stage, when the crystal nucleus is extremely small. However, twinned overgrowth may also be observed forming on a nucleus of relatively-large size\cite{donnelly1967,carstens1968}. In both of these cases, atomistic simulation suggests itself as the ideal avenue to unravel the nature of twin nucleation and growth. 

Here we report molecular dynamics simulations of Ge recrystallization from 
 amorphous substructures at high homologous temperatures (fraction of the melting temperature 
 $T_m$). Our simulations are framed within the study of amorphization-crystallization ($a$$\rightarrow$$c$) processes in laser spot heating of GST-based\footnote{Ge-Sb-Te} phase-changing materials (PCMs)\cite{Nikolova2010,Nikolova2013}. In GST materials, the main role of Ge is to accelerate the recrystallization process, and thus here we study pure Ge as the point of reference for Ge-based PCMs \cite{Wuttig2012}. We start from a critical nucleus of crystalline Ge ($c$-Ge) embedded in an amorphous ($a$-Ge) medium at an initial temperature of $T_0=1100$ K, which is representative of the conditions found within the laser spot. We find that grain growth is characterized by the formation of intricate twin networks facilitated by near zero surface and stacking fault energies\cite{posselt2009}. We accompany our simulations with a full thermodynamic analysis to explain the mechanisms behind the observed behavior.
 
\section{\label{sec:simul}Simulation details}
\subsection{Molecular dynamics simulations}
\label{subsec:md}
We use a Stillinger-Weber potential parameterized by Posselt and Gabriel for Ge\cite{posselt2009}, which reproduces the experimental values for the cohesive energy and lattice constant for the diamond-cubic structure, and yields reasonable values for the energetics of other crystalline phases and the structure of the liquid \cite{}. 
Our simulations are run in the isobaric-isenthalpic ensemble $NpH$ --where $N$ is the number of particles, $p$ is the pressure and $H$ is the enthalpy-- using periodic boundary conditions in three dimensions. The $NpH$ ensemble was chosen to allow for local temperature increases due to latent heat release during the crystallization process. To simulate the effect of laser spot heating, the system is equilibrated to an initial temperature of
$T_0=1100$ K, which is approximately the temperature at the center of the spot\cite{Nikolova2013}. The starting configuration is generated from a perfect diamond structure with lattice constant $a_0 = 0.5654$ nm, corresponding to the value at $1100$ K, oriented along the $[100]$, $[011]$, and $[0\bar{1}1]$ directions. The simulation comprises $62a_1\times 44a_2\times 44a_3$ supercell ($\approx$35 nm per side, where $a_1=a_0$, $a_2=a_3=a_0\sqrt{2}$)  containing 1,920,512 atoms. The central spherical region of radius of 2 nm (the nucleus) is then kept frozen, while the outer region is melted by fixing the temperature at 3000 K using a Langevin thermostat during 100 ps and then quenched. At the same time the $NpH$ ensemble maintains zero pressure globally. Finally, the entire system (the $a$-Ge block containing the $c$-Ge nucleus) is further equilibrated at 1100 K during an additional 10 ps. The thermostat is then turned off during the subsequent crystallization simulations. 
We have shown that this results in a glass transition temperature of approximately 810 K, in good agreement with laboratory experiments for Ge\cite{Angell2000}. The procedure just described was used by \citet{Reina2014} to generate planar $a/c$ interfaces to calculate free energies and interface mobilities.

\subsection{Calculation of the critical nucleus size}
\label{sec:critical}
\begin{figure}[h]
\centering
        \includegraphics[width=1.0\linewidth]{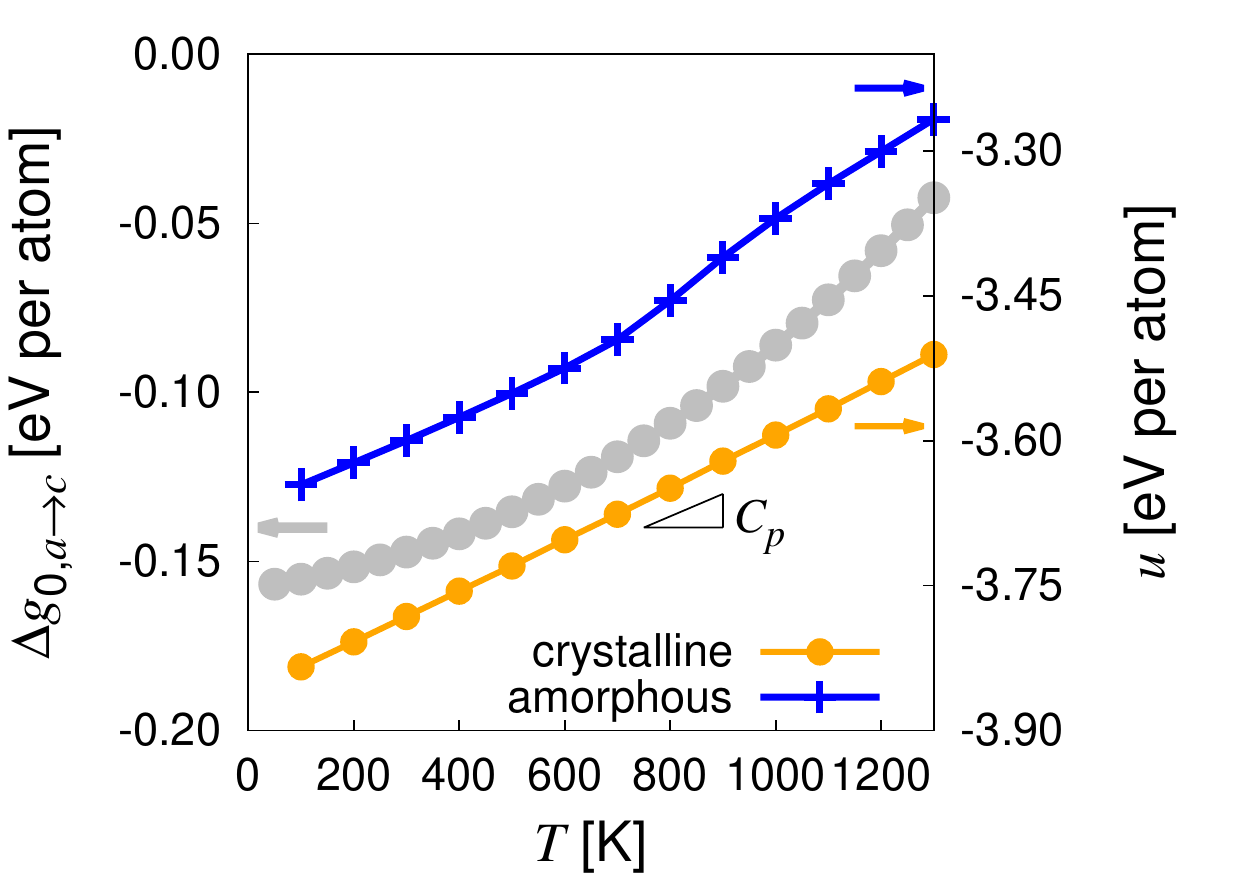}
	   \caption{\label{energies} Temperature dependence of $\Delta g_{0,a\rightarrow c}$ expressed on a per atom basis (referred to the left vertical axis). Also shown are the internal energies $u$ of both amorphous and crystalline Ge as a function of temperature (right vertical axis). The heat capacity $C_p$ is 
	   calculated from the slope of $u(T)$, which for $c$-Ge results in a value of $\approx2.69\times10^{-4}$ eV$\cdot$K$^{-1}$ per atom ($\approx0.36$ J$\cdot$g$^{-1}$$\cdot$K$^{-1}$), in excellent agreement with experimental measurements \cite{okhotin}.}
\end{figure}

In classical nucleation theory, the critical nucleus size $r^{\ast}$ is governed by the balance between the volumetric and interfacial driving forces expressed, respectively, as the derivative of the net free energy release $\Delta G_{0,a\rightarrow c}$ and a surface energy penalty $\Delta G_s$ with respect to the radius of the nucleus. 
In principle, this balance must also account for the expansion of Ge upon crystallization, which 
is approximately 10\% less dense than its amorphous counterpart (cf.~Figure 3b in \citet{Reina2014}) at $p=0$ and $T_0=1100$ K.
However, the procedure detailed in the previous section to seed an amorphous matrix with crystalline grains removes any differential strains by construction. This allows us to write: 
\begin{equation}
\Delta G=\Delta G_{0,a\rightarrow c}+\Delta G_s=\frac{4\pi r^3}{3}\Delta g_{0,a\rightarrow c}+4\pi r^2\gamma{\color{blue},}
\label{delta}
\end{equation}
where $\Delta g_{0,a\rightarrow c}$ is the volumetric free energy density at zero pressure, and $\gamma$ is the surface (free) energy density, which is orientation dependent: $\gamma\equiv\gamma(\theta)$, with $\theta$ representing the surface normal with respect to the crystal orientation. 
We have calculated the atomic free energy densities $g_{0,a}$ and $g_{0,c}$ of the amorphous and crystalline phases using thermodynamic integration, see \citet{Reina2014} for details. The variation of $\Delta g_{0,a\rightarrow c}=g_{0,a}-g_{0,c}$ with temperature in units of energy per atom is provided in Figure \ref{energies}. When this difference is zero, there is phase coexistence, which by definition occurs at the melting point, here $T_m=1350$ K. The driving force per unit volume for the $a$$\rightarrow$$c$ transformation at zero pressure is readily obtained as: $\Delta g_{0,a\rightarrow c}=\rho_c\Delta g_{0,a\rightarrow c}$. From the figure, at $T_0=1100$ K, $\Delta g_{0,a\rightarrow c}=-0.073$ eV per atom, while the atomic density of the crystalline phase at the same temperature is $\rho_c=4.36\times10^{28}$ m$^{-3}$ (after \citet{Reina2014}). From this, $\Delta g_{0,a\rightarrow c}\approx -5.07\times10^8$ J$\cdot$m$^{-3}$.

\begin{figure}[h]
\center
\includegraphics[width=\linewidth]{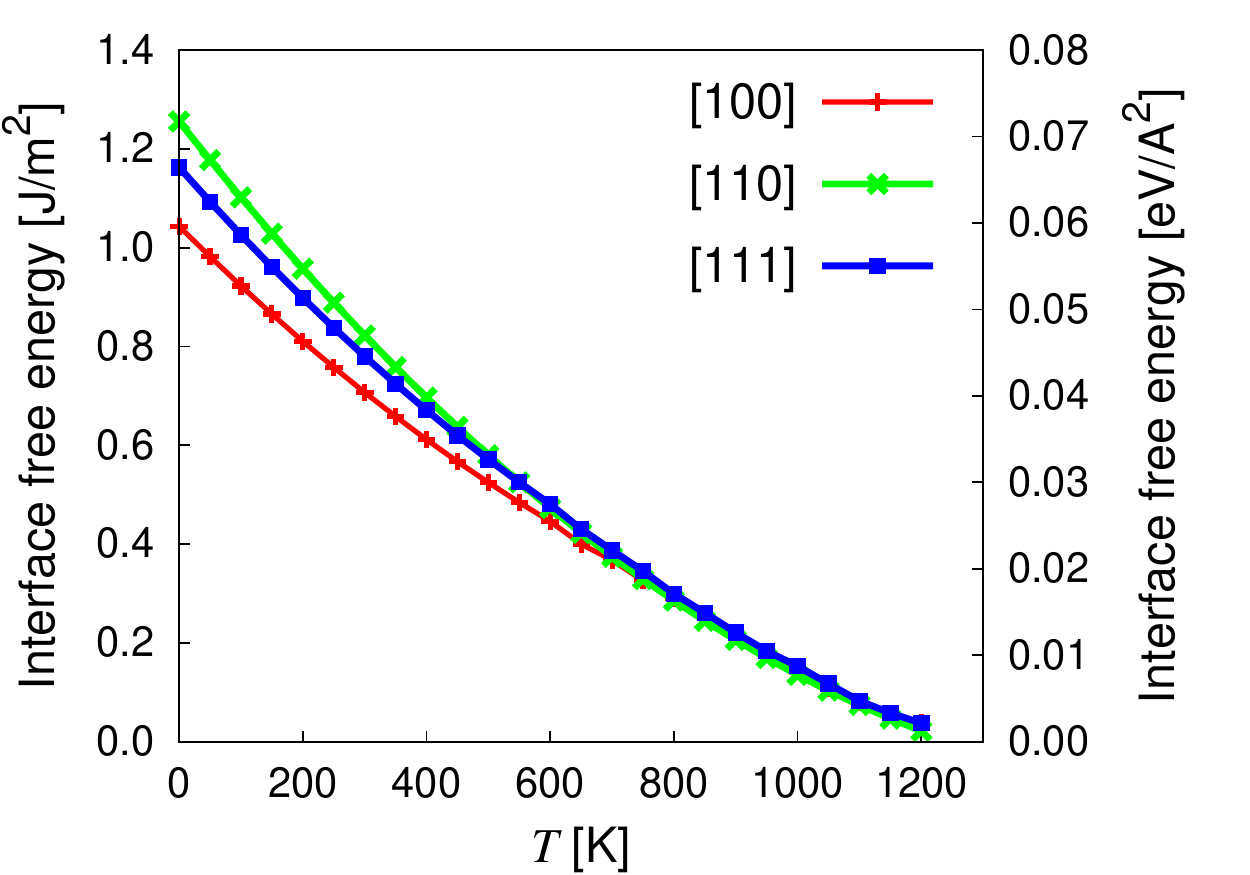}
\caption{\label{surface} Surface free energy as a function of temperature for the three surface normals representing the vertices of the standard triangle.}
\end{figure}

As mentioned earlier, $\Delta G_s$ is orientation dependent. However, crystalline Ge displays cubic symmetry, which allows us to reduce the orientation space to that contained in the standard stereographic triangle whose vertices in the first octant are the intersects of the unit sphere with the $[001]$, $[110]$, and $[111]$ directions\cite{Reina2014}. Thus, we restrict our study of the orientation dependence of $\gamma(\theta)$ to those three orientations\footnote{Suitable interpolation schemes within the standard triangle can be adopted for a general orientation $\theta$.}. The interfacial free energies are shown in Figure \ref{surface} as a function of temperature, where a surface orientation anisotropy can be clearly distinguished at low temperatures. At $T_0$ however, this anisotropy is smeared out by the high thermal diffusivity of the amorphous phase above the glass transition temperature, cf.~Figure 2 in \citet{Reina2014}, and we find an orientation-independent value of $\gamma\approx0.08$ J$\cdot$m$^{-2}$. 

The critical grain size is found by minimizing eq.\ \eqref{delta}: $d\Delta G/dr=0$, which results in
\begin{equation}
r^{\ast}=-\frac{2\gamma}{\Delta g_{0,a\rightarrow c}}{\color{blue}.}
\end{equation}
Replacing $\gamma$ and $\Delta g_{0,a\rightarrow c}$
 for their respective values, we obtain that $r^{\ast}\approx0.32$ nm. This value is approximately 55\% 
  of the magnitude of the lattice constant $a_0$ and suggests stable crystalline grains with only a handful of atoms in them, probably implying that very localized fluctuations suffice to produce crystal growth seeds.

\subsection{{\label{inter}}Interface mobility}
\begin{figure}[h]
\center
\includegraphics[width=\linewidth]{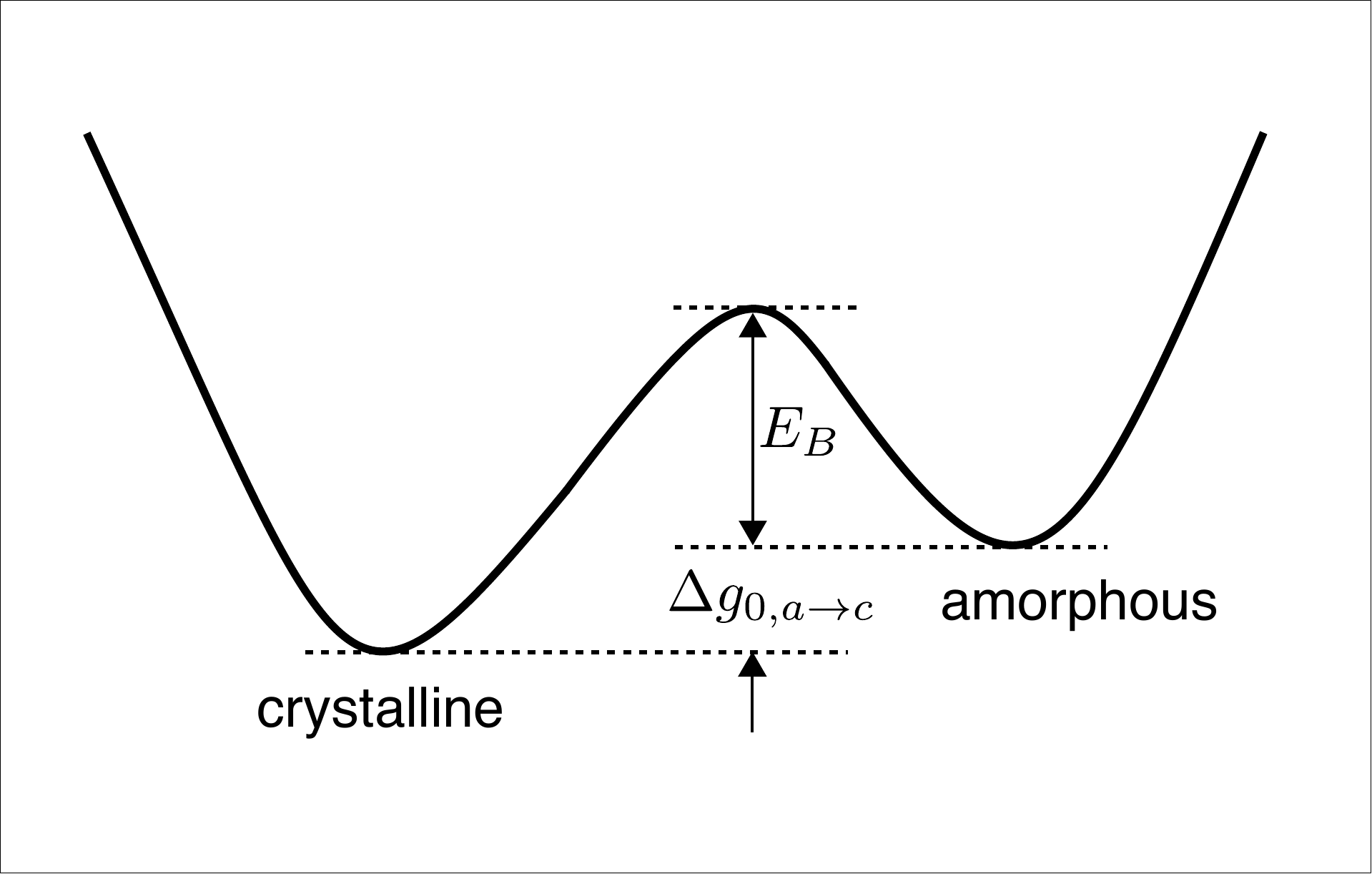}
\caption{\label{scheme} (Free) energy landscape governing the crystallization process.}
\end{figure}

\begin{figure}[h]
\center
\includegraphics[width=\linewidth]{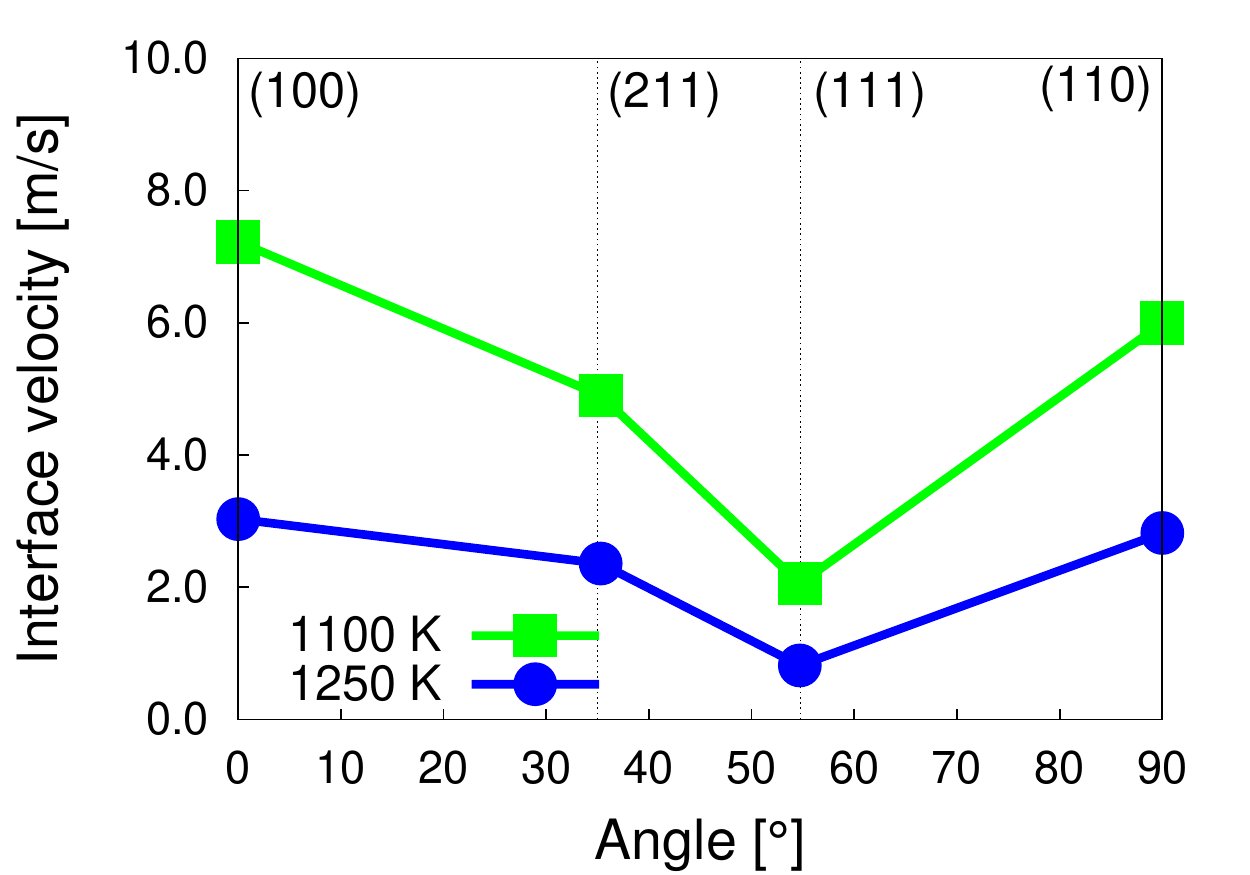}
\caption{\label{intvel} Interface velocity as a function of surface orientation at the initial (1100 K) and final (1250 K) simulation temperatures.}
\end{figure}

The growth rate of the crystalline phase at the expense of the amorphous phase is governed by the schematic energy landscape shown in Figure \ref{scheme}. The excess atomic flux $a$$\rightarrow$$c$ relative to $c$$\rightarrow$$a$ transitions is governed by $\Delta g_{0,a\rightarrow c}$ (shown in the figure) and results in net interface velocity $v(T)$ and grain growth. Mathematically, this can be expressed to first order as\citep{bourgoin1984solid,liu2014multi}: 
\begin{equation}
v(T)=v_0\exp\left(-\beta E_B\right)\left(1-\exp\left(-\beta\Delta g_{0,a\rightarrow c}\right)\right)
\end{equation}
where $v_0$ is a prefactor and $E_B$ is an activation energy for the transformation (shown in the figure).  $E_B$ represents the energy for the detachment/reattachment process, which is diffusive in nature. We have devised a special procedure to calculate $E_B$, for which a value of 0.42
 eV was obtained for the $[100]$ orientation\cite{Reina2014} at 0 K. A low value of $E_B$ may result in faster growth speeds at low temperatures, but it also results in faster detachment ($a$$\leftarrow$$c$) at higher temperatures, where the difference of free energies decreases, resulting in lower effective growth speeds. 
The interface velocities can also be obtained by direct atomistic simulation as described by \citet{Reina2014}.  Figure \ref{intvel} shows results for four distinct orientations at $T_0$ with the $\langle111\rangle$ being the slowest one, which ultimately controls grain growth.

\section{{\label{results}}Simulation results}

Next we present the simulations of supercritical grain growth at $T_0$. While the critical radius calculated earlier suggests a very small stability threshold, we have found that --in practice-- a minimum radius of 2 nm was needed to have positive grain growth on the timescales captured in the MD simulations. The discrepancy can be attributed to a number of factors, chief of which is the magnitude and frequency of 
thermal fluctuations at these high temperatures and small volumes, which lead to low signal-to-noise ratios in terms of the stable critical size. Other factors such as nonsphericity, and finite size effects, may also play a non-negligible role. Consequently, in the following we show results of 2-nm radius supercritical $c$-Ge nuclei in an amorphous medium.

\subsection{Three-dimensional growth of critical grains}

\begin{figure}[h]
\centering
	\subfigure[]{
		\boxed{
			\includegraphics[width=0.65\linewidth]{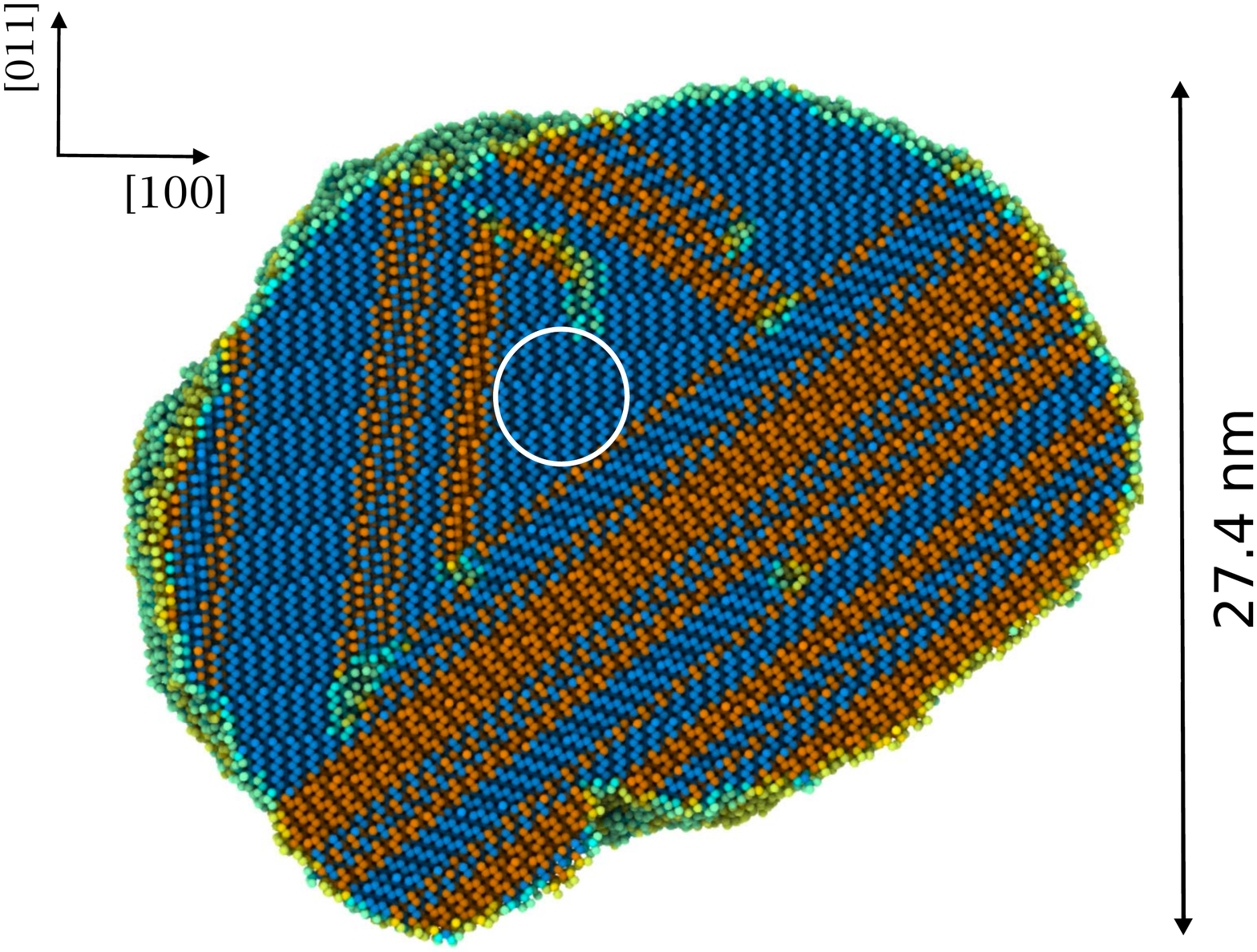}}
    	\label{subfig1}}
    	\subfigure[]{
        \boxed{
        \includegraphics[width=0.65\linewidth]{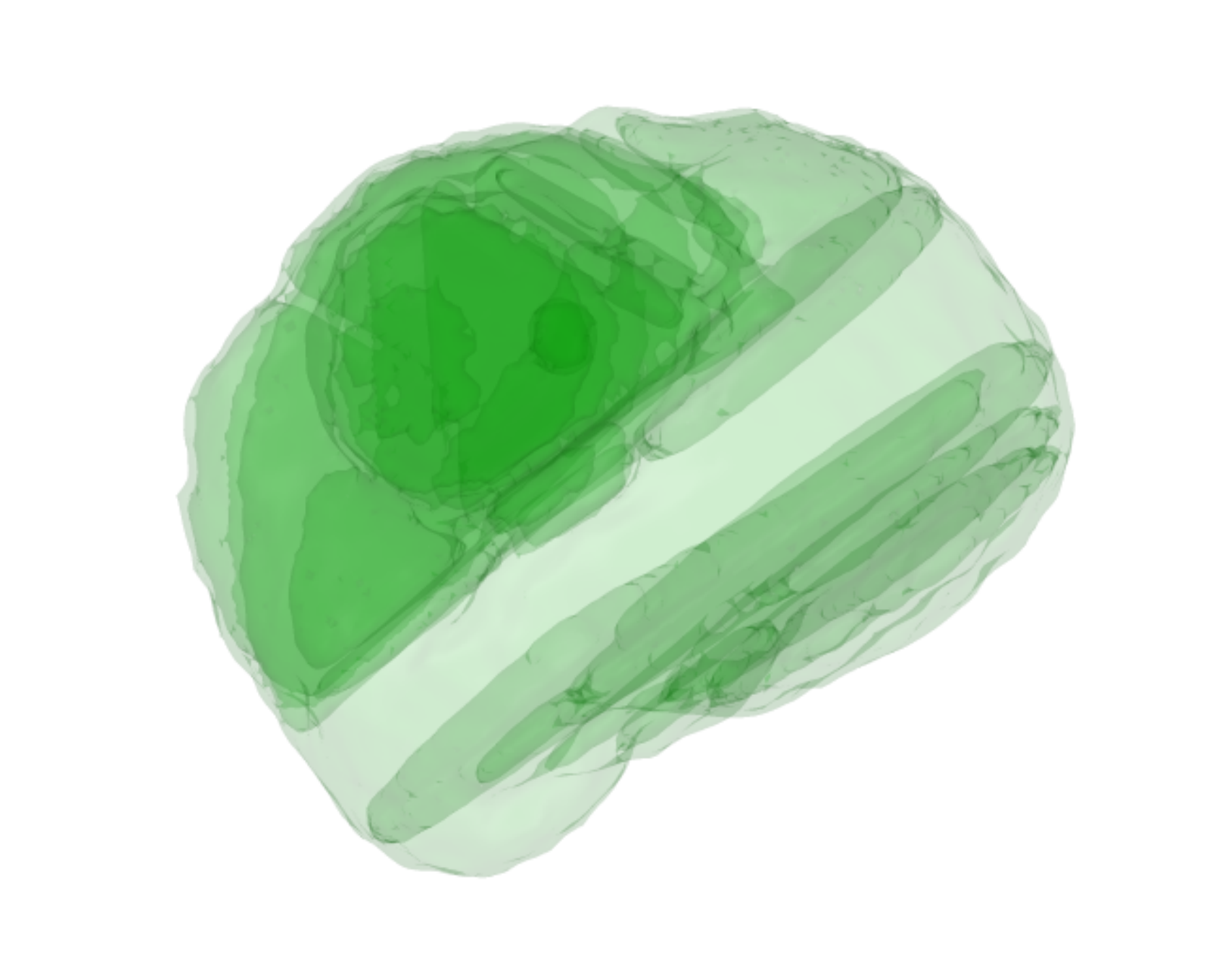}}
    	\label{subfig2}}
	   \caption{\label{snapshot} Structure of the crystalline grain after 9.1 ns of growth in the $NpH$ ensemble. (a) Atomistic structure: Only atoms possessing an ordered crystallographic structure are shown. Dark blue and orange spheres represent atoms with diamond cubic and twin plate structure, respectively. Light blue spheres represent atoms belonging to the amorphous/crystalline interface and dislocation cores. The circular region in the center of the image indicates the extent of the critical grain at the beginning of the simulation. (b) Through-thickness view of the nanotwin structure corresponding to (a).}
\end{figure}

In a 3D (spherical) nucleus, in principle all growth orientations are sampled, which means that in materials with sizable interface energy anisotropies and/or interface velocities, some growth directions will be preferred over others. A complete animation of the grain growth process starting from (super)critical nuclei is provided in the Supporting Information. A snapshot of the simulation at the point of maximum growth --which occurs 9.1 ns after the system is equilibrated at $T_0$-- is shown in Figure \ref{snapshot}. Atoms in the image are colored using structure analysis as implemented in the OVITO visualization package \cite{ovito}, which assigns dark blue to atoms with diamond cubic structure and orange to atoms with hexagonal diamond crystal structure.

\begin{figure}[h]
\center
\includegraphics[width=\linewidth]{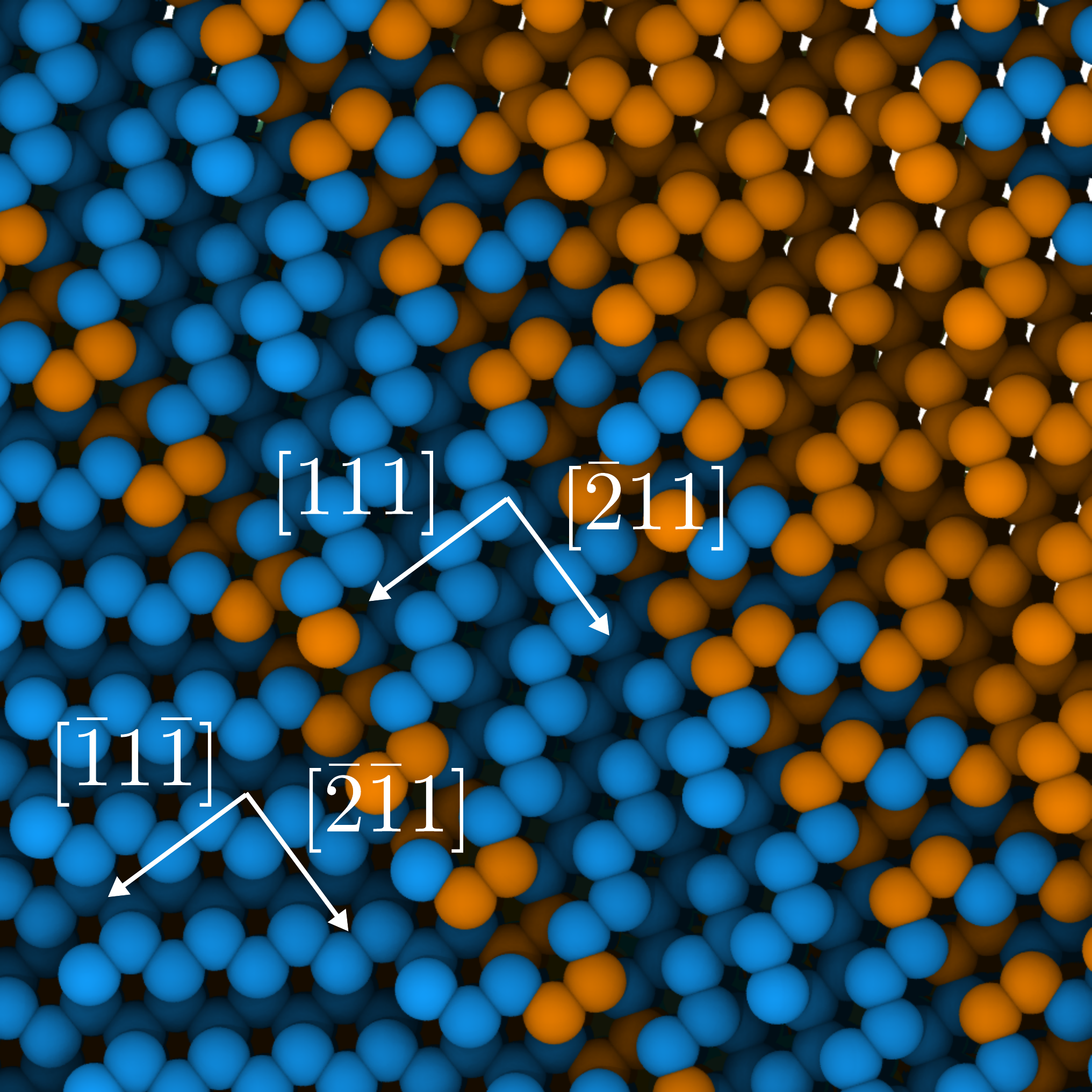}
\caption{\label{zoom} $[0\bar{1}1]$ view of the atomistic configuration of a twinned region.}
\end{figure}

Closer examination of the atoms with hexagonal diamond structure reveals that they belong to $\langle111\rangle$ twin ($\Sigma3$) boundaries, typical of the diamond cubic lattice structure. Figure \ref{zoom} shows a region around one such boundary in local detail, where the mirror symmetry characteristic of twin plates can be clearly identified. 
Moreover, using boundary analysis available in OVITO, we have examined the atomistic structures in Figures\ \ref{subfig1} and \ref{zoom} and find the emergence of a network of twinned regions, as showcased in Figure \ref{subfig2}. The figure shows a through-thickness view of the entire grain at the exact same time as Fig.\ \ref{subfig1}. As depicted, in the lower half of the grain twins are elongated along $\langle211\rangle$ directions, while in the upper half a three-dimensional arrangement is formed.

\begin{figure}[h]
\center
\includegraphics[width=\linewidth]{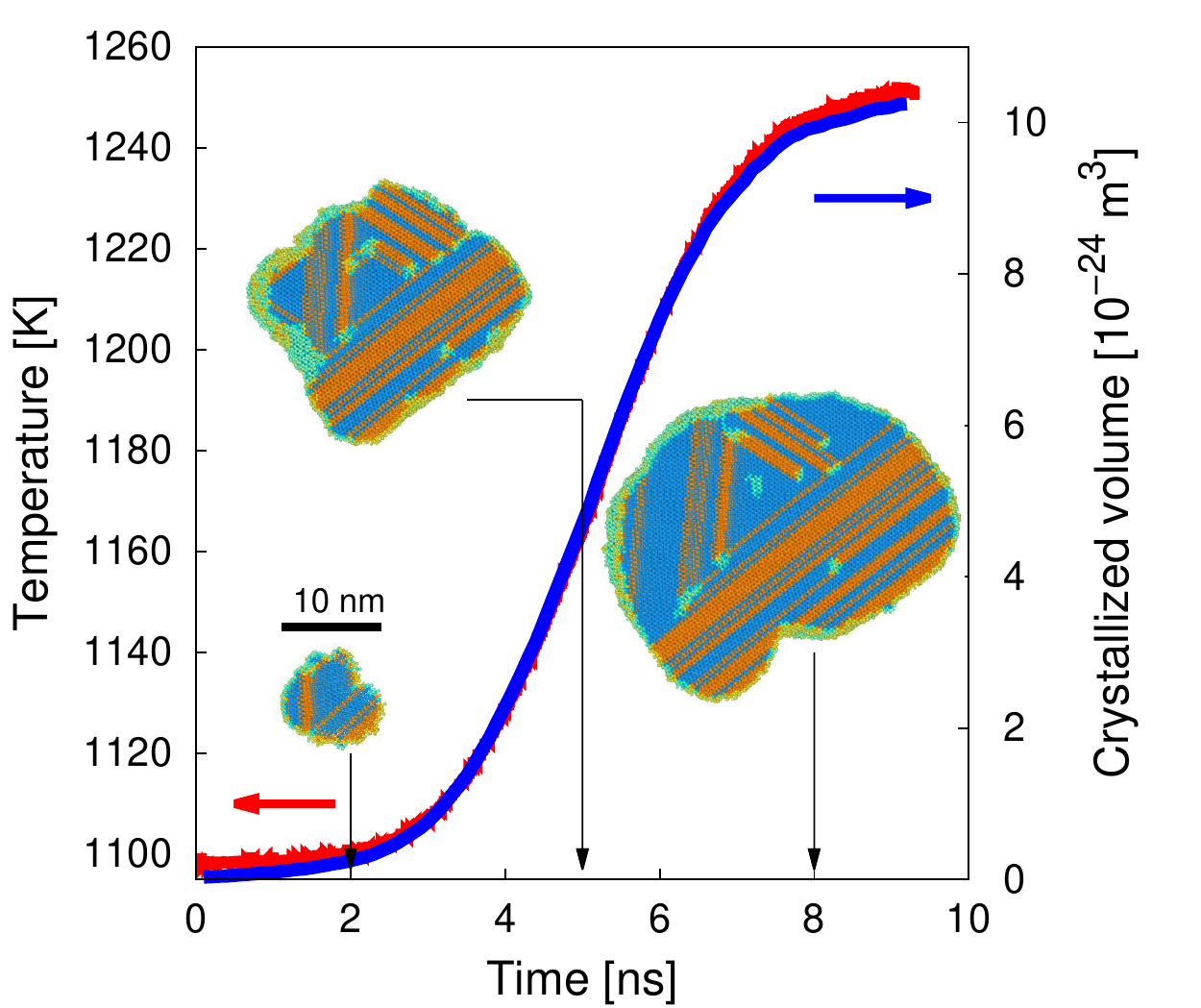}
\caption{\label{temp} Evolution of the temperature with time during growth of a critical nucleus under $Nph$ simulations conditions. The crystallized volume is also shown as a function of time. Selected snapshots of the grain structure corresponding to three distinct instants (2, 5, and 8 ns) are shown.}
\end{figure}

As discussed earlier, grain growth is controlled by the magnitude of the driving force and the interface mobility. Both of these quantities are temperature dependent. The exothermic nature of the $a$$\rightarrow$$c$ reaction (i.e.~internal energy density difference $\Delta u_{a\rightarrow c}<0$) results in a local energy deposition that increases the global system temperature. Thus, the temperature of the system correlates directly with the volume of material transformed. Such correlation is clearly visible in Figure \ref{temp}, where both quantities are exactly proportional to each other with a proportionality constant of $\approx8.01\times10^{-27}$ m$^3$ K$^{-1}$. 
The temperature is seen to increase from $T_0$ to a final value of approximately 1250 K. This effectively arrests the growth process, as dictated by the sharp decrease in mobility at such temperature (cf.\ Figure\ \ref{intvel}). We remark that this arrest is partially an artifact of the simulations which limits unrestricted heat flow due to periodic boundary effects. 

\subsection{Analysis of growth of $a/c$ bicrystals}

The appearance of twins during the growth stage of critical Ge grains may obey the energetics of two different scenarios. On the one hand, there is ample evidence in the literature that in Ge twinning emerges if the growth direction deviates appreciably from the preferred growth direction, understood as that which results in the lowest interfacial energy. Twins would then appear to alter the internal crystal orientation and bring it closer to the preferred one \cite{billing1954,kohn1958}. However, other works have pointed out that if a new facial orientation was the only advantage gained by twinning, twinned crystals should not be much larger than twice the size of a single crystal, something at odds with observations of twinned crystals being more than ten times as large as untwinned ones \cite{carstens1968}. This hypothesis is strongly weakened by the lack of a noticeable surface energy anisotropy at 1100 K according to our calculations (cf.\ Figure\ \ref{surface}).

The alternative scenario is that twins are a manifestation of a growth mode that relies on the indistinct formation of ordered atomic layers with the correct stacking sequence and stacking faults. 
This is the same growth mode observed under the so-called \emph{solid-phase epitaxial recrystallization} (SPER) process of $\{111\}$-oriented crystals, as well as by \emph{liquid} epitaxial growth of crystals with the same orientation via chemo-physical vapor deposition. There is ample evidence of twin formation in the literature for both of these processes in Ge, particularly at high temperatures \cite{cor2008,darby2012,epitaxy-book}. 
This mechanism is controlled by the value of the stacking fault energy $\gamma_{\rm SF}$\footnote{Twin boundaries correspond to `half' a stacking fault and so the twinning propensity correlates directly with the stacking fault energy}, which ranges between 0.07 and 0.09 J$\cdot$m$^{-2}$ according to several measurements\cite{1963ApPhL,gomez1975,denteneer1987}. By contrast, the interatomic potential employed in our simulations predicts zero stacking fault energy\cite{posselt2009}. Evidently then, the model for Ge employed here offers no impediment to the favorable formation of epitaxial twins. However, while at low temperatures this might clearly result in an overestimation of the volume fraction of stacking faults and/or twins when conditions are conducive to their formation \cite{biswas2012,Nguyen2013}, it is reasonable to assume that values of $\gamma_{\rm SF}$ on the order of the experimentally-measured ones result in zero effective stacking fault energy at a temperature of 1100 K via thermal softening. 

\begin{figure}[h]
\center
\includegraphics[width=0.8\linewidth]{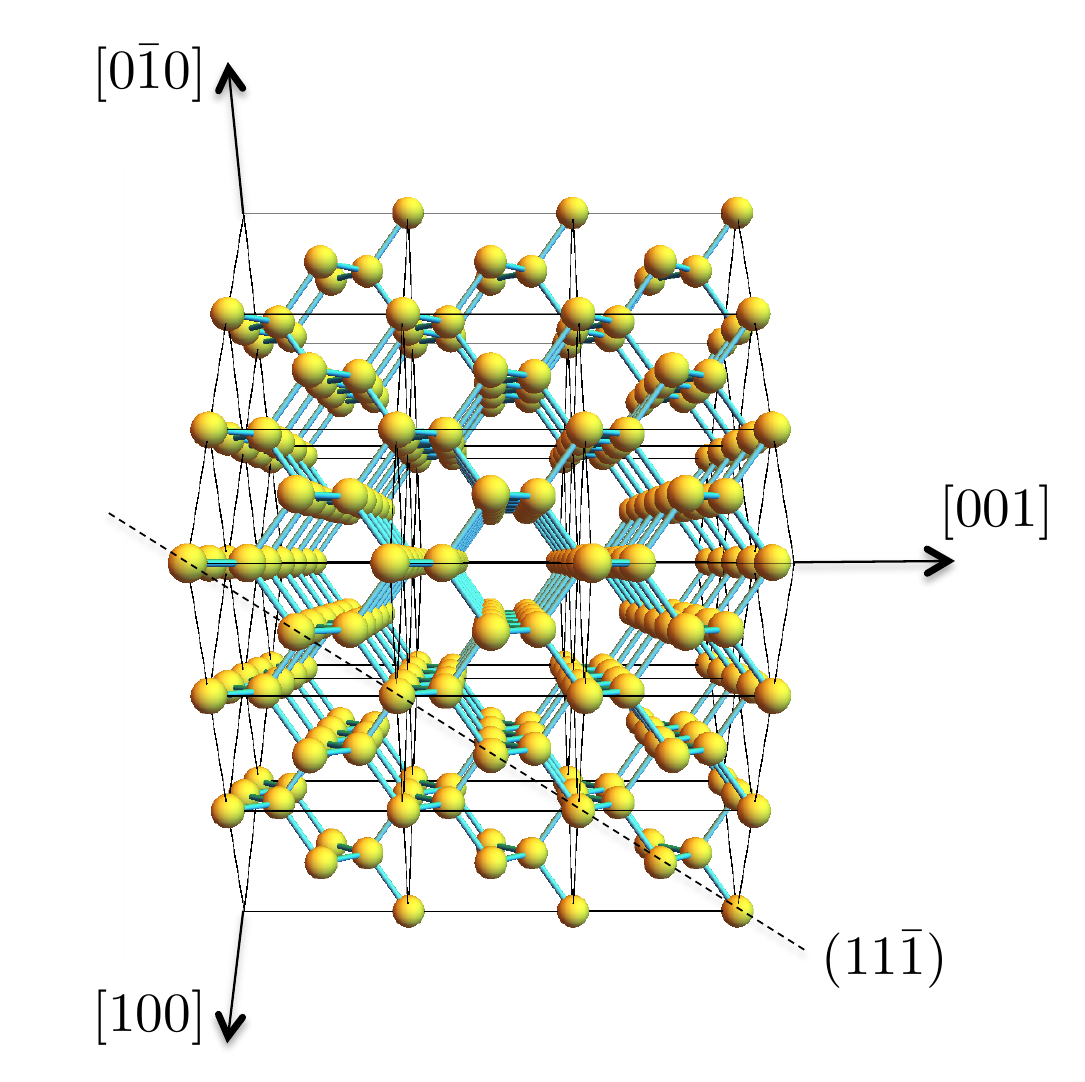}
\caption{Schematic view along the $[1\bar{1}0]$ direction of the Ge diamond cubic lattice, with the $(11\bar{1})$ plane highlighted. Image obtained with \texttt{Wolfram CDF Player}\cite{WolframAlpha}.}
\label{red}
\end{figure}

To ascertain which mechanism is responsible for the observed formation of nanotwin networks, next we carry out MD simulations of $a/c$ bi-crystals at $T_0$ oriented along three selected directions: $[111]$ (low mobility, cf.\ Figures\ \ref{surface} and \ref{intvel}), and $[100]$ and $[110]$ (high mobility). These are qualitatively similar to other simulations of the SPER process using atomistic methods\cite{george1993,akis2007,bragado2012}. The three surface orientations simulated here are schematically shown in Figure\ \ref{red} relative to a $[1 \bar{1}0]$ view of the Ge diamond cubic lattice.

\subsubsection{$[111]$ amorphous/crystalline bi-crystals}
\label{111}
\begin{figure}[h]
\center
\includegraphics[width=\linewidth]{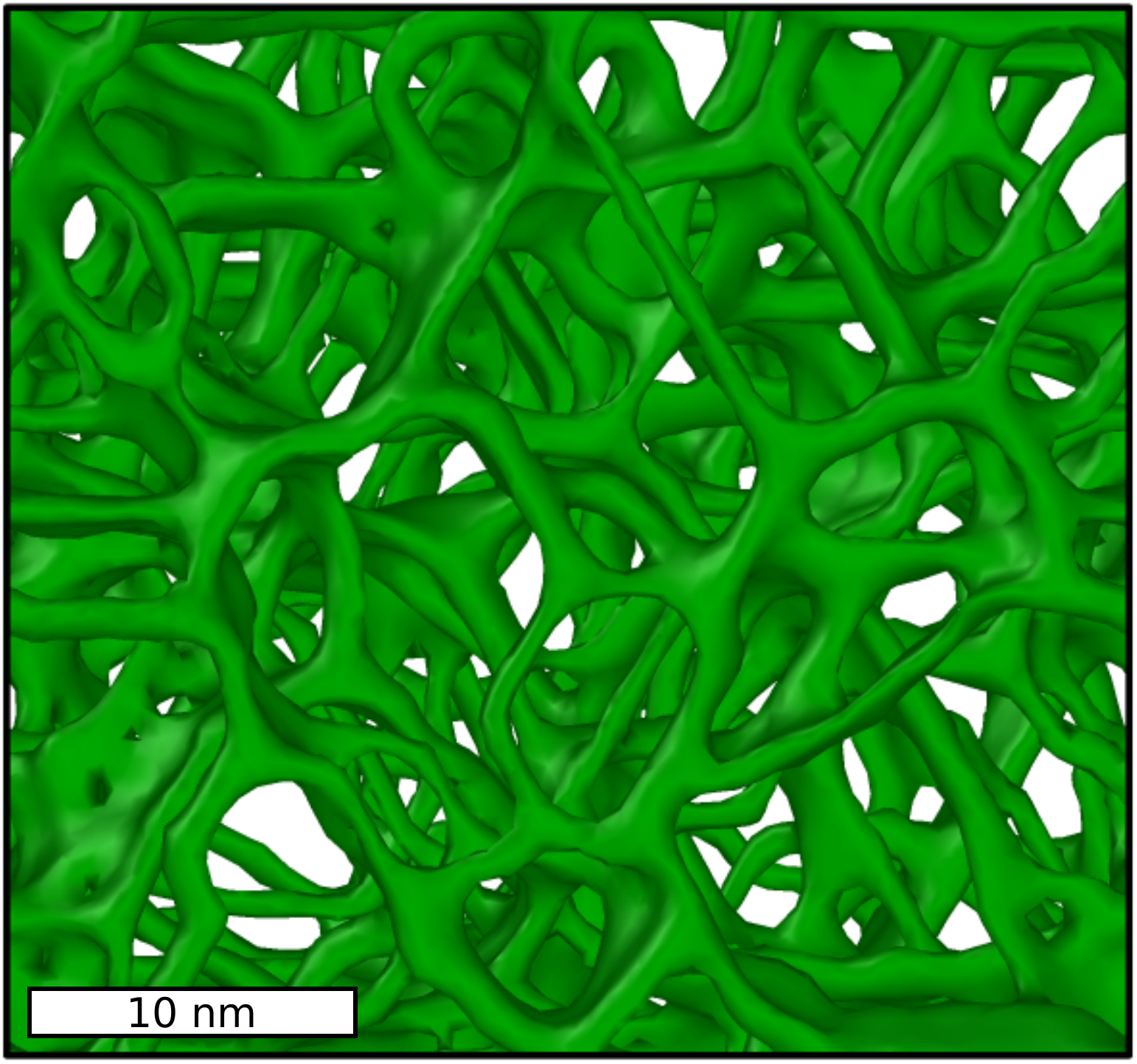}
\caption{\label{twin111} Twinning network formed from the growth of a crystalline Ge half-crystal along the $[111]$ direction.}
\end{figure}

The starting microstructures ($a/c$ bi-crystals) are generated in the manner described by \citet{Reina2014}. The computational cell has dimensions of $20.8\times19.4\times39.2$ nm containing 698,880 atoms. The system is again equilibrated at 1100 K and let to evolve in the $NpH$ ensemble. Two animations illustrating the process are provided in the Supporting Information. Growth of the crystalline phase proceeds via the formation of an intricate twin network, an image of which is shown in Figure \ref{twin111}. Twin boundaries are shown as green-colored surfaces, and are seen to form a quasi-hexagonal network as demanded by the topological structure of a set of interconnected $\{111\}$ surfaces. The structures are reminiscent of coral-like porous networks in synthesized ceramic materials \cite{caruso1998}. 

\subsubsection{$[100]$ and $[110]$ amorphous/crystalline bi-crystals}

The size of the computational cells employed to study growth along the [100] and [110] directions was $20.4\times20.4\times39.6$ nm with 725,760 atoms, and $20.4\times20.8\times40.0$ nm with 748,800 atoms, respectively.
As shown in the corresponding animations (Supporting Information), crystallization along these directions results in growth of a homogeneous Ge crystal, forming essentially no defects. Here, atom rearrangements from the amorphous into the crystalline phase occurs by forming atomic planes with the correct stacking sequence.

\subsubsection{XRD analysis}

\begin{figure}[h]
\center
\includegraphics[width=\linewidth]{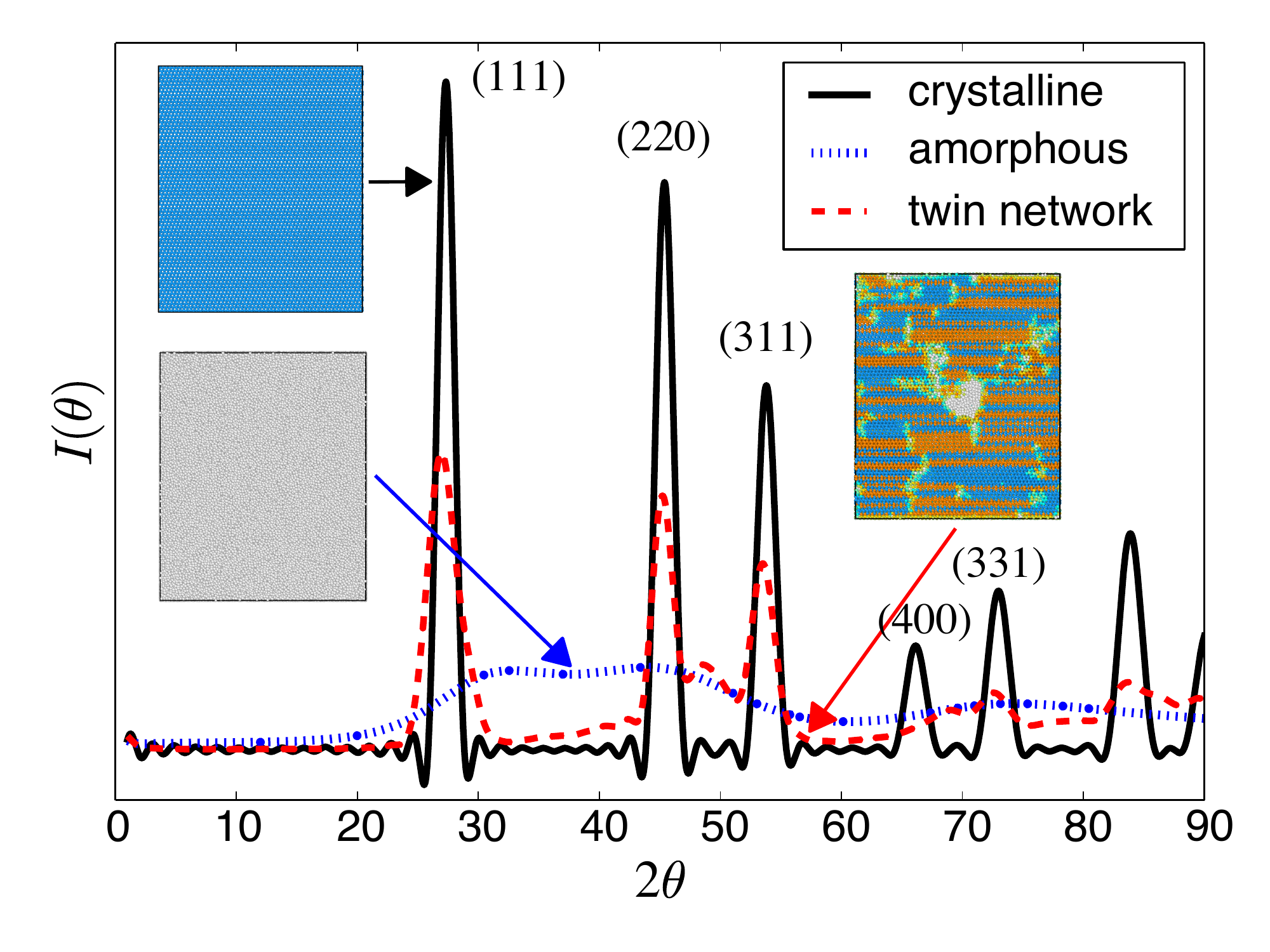}
\caption{\label{diffraction}Diffraction patterns for amorphous, perfect crystal and twinned Germanium. The atomic structures were minimized before being analized with \texttt{Debyer}\cite{debyer} considering a X-ray source with wave length of 1.542 \AA.}
\end{figure}

An important part of the analysis of the simulations is their experimental validation. Ge crystals can be examined by a variety of sources, from transmission electron microscopy (TEM), to Raman spectroscopy (RS), and X-ray diffraction analysis (XRD).
However, twin boundaries are not sources of strain and are thus difficult to detect via conventional TEM analysis. In contrast, they act as scattering agents to X-rays and do leave an imprint on diffraction patterns. Therefore, we have calculated the equivalent XRD signature for pure crystalline and amorphous samples, as well as for the multitwinned structure shown in Fig.\ \ref{twin111}, using the \texttt{Debyer} code\cite{debyer} considering a X-ray source with wave length of 1.542~\AA\ at 0 K. The resulting pattern is shown in Figure \ref{diffraction}, where the intensity peaks represent the different scattering directions. The figure reveals clear differences in the footprints of the three structures considered, namely, no structure for the amorphous system, well marked peaks for the ideal crystal, and softened peaks for the twinned crystal.
The XRD pattern showed in the Figure  is in excellent agreement with experimental results for pure crystalline Ge but only in modest agreement for amorphous Ge \cite{choco2009}. This may indicate that the generated amorphous structures may not be fully optimized in terms of their atomic configuration, likely a result of using unphysically-high heating and cooling rates to entrap a liquid structure into a disordered solid. With regard to multitwinned structures, Fig.\ \ref{diffraction} provides a pathway for their detection in future experiments.

\section{Discussion and conclusions}

Although it is clear from the literature that twinned Ge crystals may emerge during crystallization at low temperatures to favor low energy interface orientations (and thus decrease the critical nucleus size), our simulations conclusively show that the origin of the twinning network observed during crystallization of Ge grains from amorphous structures at $T_0$ lies in the energetic degeneracy observed for the stacking sequence of $\{111\}$ planes. Although this effect is favored by construction in our simulations (due to a zero stacking fault energy predicted by our atomic model), the overall effect of $\gamma_{SF}$ in materials such as Ge at these high temperatures is likely to be negligible in any case. The result is the spontaneous formations of multiply twinned structures along each of the three equivalent $\langle111\rangle$ directions.

The other notable observation is that grain growth at 1100 K is controlled by low interface mobilities. This together with a small critical radius for stable crystalline nuclei, suggests the development of nano crystalline or very fine-grained structures, as is indeed the case experimentally. The reason for this is that 1100 K is near the tipping point where the mobility sharply decreases from its maximum value. This is compounded by latent heat deposition released during the exothermic $a$$\rightarrow$$c$ process, which increases the temperature beyond that tipping point. The phenomenon where crystallization is fueled by the intrinsic latent heat release is well known and referred to as \emph{explosive} crystallization\cite{leamy1980,grigor2006,Nikolova2014}. For this, however, an increase in temperature should result in growth acceleration by a surge in interface velocity, which is not the case in the temperature regime considered here. In all, high nucleation rates due to small critical radii, slow mobilities due to high temperatures, plus high-twinning propensities result in the notoriously fine-grained nanostructures reported for laser-induced Ge crystallization \cite{grigor2006,kuo2008,Nikolova2013,Nikolova2014}. We have recently proposed a thermodynamically-consistent phase field model to predict these microstructures \cite{Reina2014}. However, intrinsic twinning was not a feature of those simulations and we believe that the present atomistic simulations provide a new piece of physics that must be incorporated into such higher-level models.

Twin boundaries may also act as scattering agents for elastic and electromagnetic waves, and may impact the value of fundamental constants such as the thermal conductivity or electric susceptibility. Indeed, it has been observed that the appearance of twins during epitaxial growth of Si wafers resulted in faulty devices, while for other defects, such as {\it e.g.}\ extrinsic stacking faults, it did not \cite{lee2006,checos2012,china2013} (albeit perfect Ge crystals have been grown in the $[111]$ direction as well\cite{Nguyen2013}). This may be of importance in GST materials where  high contrast between amorphous and crystalline phases in terms of these properties is desired.

\begin{acknowledgement}:

We thank the DTEM group at Lawrence Livermore National Laboratory for useful suggestions and guidance. This work was partially performed under the auspices of the US Department of Energy by Lawrence Livermore National Laboratory, and Los Alamos National Laboratory. LLNL is operated by Lawrence Livermore National Security, LLC, for the National Nuclear Security Administration of the U.S. DOE, under contract DE-AC52-07NA27344. LANL is operated by Los Alamos National Security, LLC, for the National Nuclear Security Administration of the U.S. DOE, under contract DE-AC52-O6NA25396.
\end{acknowledgement}

\begin{suppinfo}

Animations referred to in the text as `Supporting Information' will be made available online.

\end{suppinfo}

\bibliography{Biblio}


\begin{tocentry}
\begin{center}
\includegraphics[width=0.35\linewidth]{111_twin_net.pdf}
\end{center}
Twinning network formed during high temperature Ge crystallization.
%
%
%

\end{tocentry}

\end{document}